\documentclass[twocolumn,pra,longbibliography]{revtex4}
\usepackage{amsmath}
\usepackage{graphicx}
\usepackage{soul,xcolor}
\setstcolor{red}
\definecolor{mycolor}{rgb}{0.1, 0.1, 0.7}
\usepackage{dsfont}
\usepackage{url}
\usepackage[colorlinks]{hyperref}

\hypersetup{
 plainpages=true,
 breaklinks=true,
 hypertexnames=false,
 pageanchor=true,
 colorlinks=true,
 linkcolor={mycolor},
 citecolor={mycolor},
 urlcolor={mycolor},
 anchorcolor={black}
 }
\newcommand{\ket}[1]{|{#1}\rangle}
\newcommand{\bra}[1]{\langle{#1}|}

\newcommand{\e}[1]{e^{#1}}

\begin{document}
\title{Ancilla mediated qubit readout and heralded entanglement between rare-earth dopant ions in crystals}
\author{Kamanasish Debnath}
\email{kamanasish.debnath@phys.au.dk}
\author{Alexander Holm Kiilerich}
\email{kiilerich@phys.au.dk}
\author{Klaus M{\o}lmer} 
\email{moelmer@phys.au.dk}
\affiliation{Center for Complex Quantum Systems, Department of Physics and Astronomy, Aarhus University,
Ny Munkegade 120, DK-8000, Aarhus C, Denmark}
\date{\today}

\begin{abstract}
Owing to their long excited state lifetimes, rare-earth ions in crystals are widely used in quantum applications. To allow optical readout of the qubit state of individual ions, we propose to dope the crystal with an additional nearby ancilla ion with a shorter radiative lifetime. We show how a Bayesian analysis exhausts the information about the state of the qubit from the optical signal of the ancilla ion. We study the effects of incoherent processes and propose ways to reduce their effect on the readout. Finally, we extend the architecture to ions residing in two remote cavities, and we show how continuous monitoring of fluorescence signals from the two ancilla ions leads to entanglement of the qubit ions. 
\end{abstract}

\maketitle

\section{Introduction}
During the past decades, impurities and defects in solid state systems such as rare-earth ions in crystals~\cite{Casabone2018,PhysRevB.77.125111,Gobron:17}, quantum dots in nanoscale semiconductors~\cite{PhysRevLett.119.143601,PhysRevA.57.120,Petta2180} and nitrogen vacancy (NV) centers in nanodiamonds~\cite{Young_2009, Kaupp2016,PhysRevLett.112.047601,PhysRevLett.97.087601} have emerged as promising platforms for quantum technologies~\cite{Weber8513}. 
The popularity of these systems can be attributed to their outstanding coherence properties~\cite{nature_coherence,nature_coherence2} and wide operating regimes~\cite{PhysRevB.74.161203, Bradac2017, PhysRevLett.103.070502}.
For instance, isolated rare-earth ions in crystals can be used for robust quantum gates~\cite{PhysRevA.69.022321} and high precision sensors~\cite{doi:10.1063/1.4751349} while ensembles of rare-earth ions are excellent candidates for realising quantum memories and collective quantum effects~\cite{PhysRevLett.114.170503, Zhong1392,arxivpreprint1,PhysRevA.100.053821}.
The strong dipole-dipole interactions between nearby ions can be used in a similar manner as the Rydberg excitation blockade mechanism~\cite{PhysRevLett.85.2208, PhysRevLett.87.037901} to implement quantum gates between closely situated ions~\cite{PhysRevA.69.022321, OHLSSON200271}.
However, their long coherence and excitation lifetimes prevent fast and reliable optical readout of the qubit states.

\begin{figure}
\centering
\includegraphics[width=0.45\textwidth]{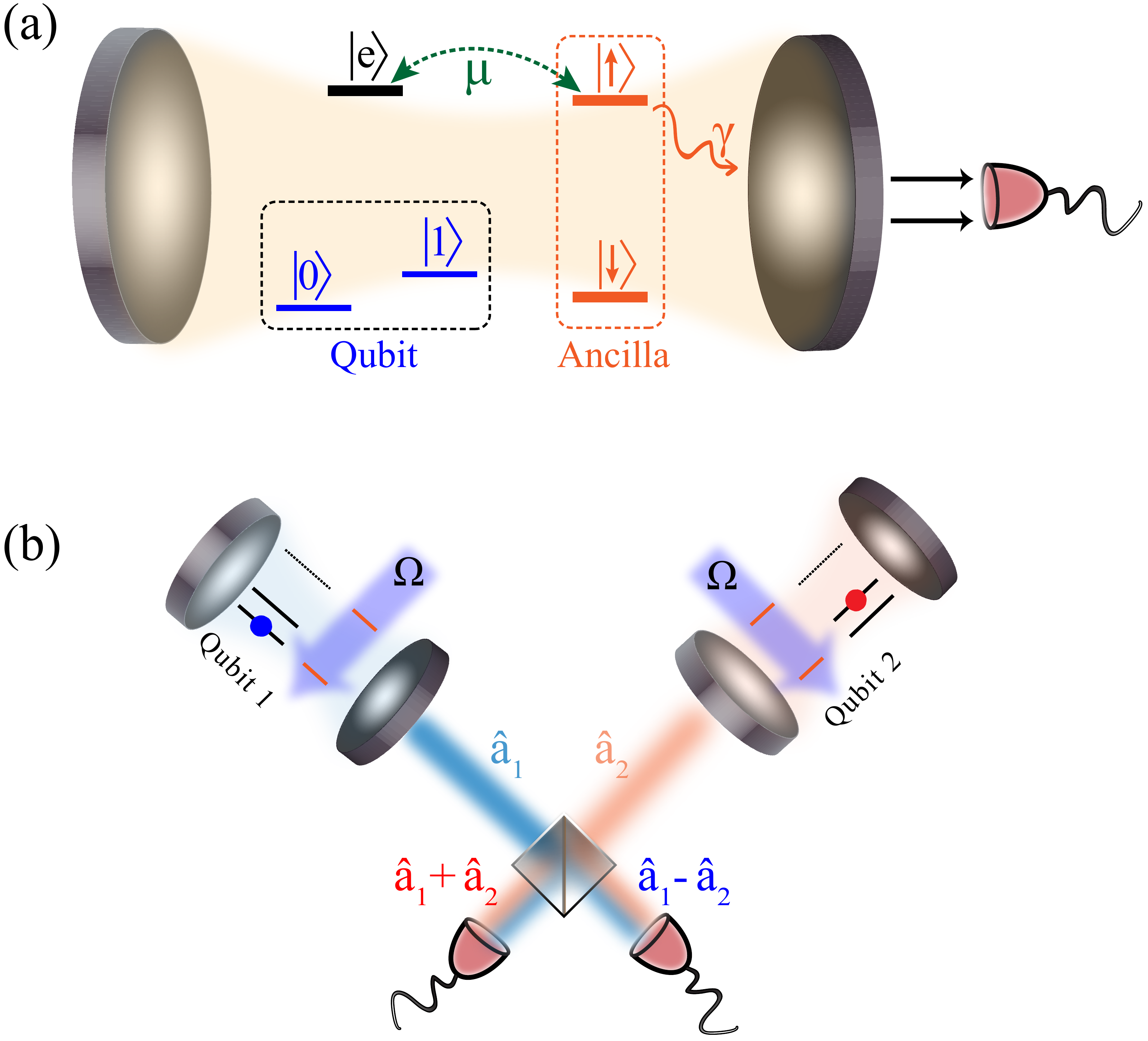}
\caption{Schematic of the ancilla based (a) readout and (b) entanglement protocols considered in this article.
(a) A qubit ion with ground states $\ket{0}$ and $\ket{1}$
is coupled to a readout ion through an excited state interaction (1).
The fluorescence signal from the readout ion is collected by a cavity mode and emitted towards an efficient photon counter,
and the signal is used to infer the qubit state.
(b) The same architecture as in (a) is used with two qubits and two readout ions to create entangled states between the two qubits by mixing the two fluorescence signals on a beamsplitter and monitoring both output channels.
}
\label{F1}
\end{figure}

In this article, we propose to separate the qubit storage and manipulation from the readout by introducing an ancilla ion. A schematic of the proposed  architecture is shown in Fig.~\ref{F1}(a). The \textit{readout} ion is resonantly coupled to a cavity mode such that its effective coupling to the quantized electromagnetic field is Purcell enhanced, thereby allowing a substantial fluorescence signal to be emitted via the cavity mode towards a photon counter.
We propose to use an architecture with two low lying qubit levels $\ket{0}$ and $\ket{1}$ and an excited level $\ket{e}$, which interacts strongly with the excited state $\ket{\uparrow}$ of the ancilla ($\hbar = 1$)
\begin{align}\label{eq:dipole}
H_{\rm dipole} = \mu \Big(\ket{e}\bra{e} \otimes \ket{\uparrow} \bra{\uparrow}\Big)
\end{align}
Depending on whether the qubit is in the excited state or not, the readout ion experiences an energy shift $\mu$, and this leads to a change in the output signal when the readout ion is subjected to a resonant continuous drive. 

Such ancilla-based quantum measurement schemes have proven beneficial in different settings~\cite{PhysRevLett.112.070502, natcom_non_dem, Schmidt749, Dutt1312,PhysRevLett.124.220501,PhysRevA.92.042710,PhysRevA.94.053830,PhysRevA.101.012309,Lachance-Quirion425}, and here we will focus our discussion on implementations in rare-earth ion systems.
Doping an inorganic crystal like YAlO$_3$, Y$_2$O$_3$ or Y$_2$SiO$_5$ with different species of rare-earth ions leads to different coherence properties due to complex interactions between the host and the dopants~\cite{doi:10.1002/zaac.201700425}. 
Hence, one can aim to engineer a hybrid system with two species of ions, where the ion with excellent coherence and lifetime properties can be used as a logical qubit, while the other, which has much faster decay rate may serve as the readout ion. 
The decay rate of the readout ion can be further enhanced by the Purcell effect, which also guarantees that the ion predominantly emits through the cavity for efficient photo detection.
Experimentally, there has been significant progress in this direction confirming the feasibility of such architectures with rare-earth ions in  crystals~\cite{PhysRevA.101.012309, SERRANO201493, SERRANO2016102}.
Alternatively, one can use a single species of ion like Eu$^{3+}$ as both qubit and readout as long as their transition frequencies are sufficiently separated. The inhomogeneous broadening in rare-earth systems ranges in the orders of a few GHz and in a small crystal with few dopant ions one can identify two closely situated ions with widely separated transition frequencies.
Recent experiments~\cite{arxivpreprintpurcell,Casabone2018} report a Purcell factor exceeding $500$, such that by making the optical cavity resonant with a single optical transition of the readout Eu$^{3+}$ ion, it experiences an effective decay rate which is $500$ times faster than that of the qubit ion.

While the integrated fluorescence signal from the readout ion may in some cases suffice to infer the state of the qubit, it is desirable to perform the readout as swiftly and precisely as possibly. This requires optimized processing of the stochastic detection signal and its temporal correlations ~\cite{PhysRevA.89.052110, PhysRevA.91.012119} and is acomplished automatically by a Bayesian analysis of the measurement signal~\cite{PhysRevA.87.032115, PhysRevA.98.022103, Sayrin_2011}. In this work we investigate the performance of such an analysis in detail.
We consider different cases and explore regimes where the Bayesian inference is clearly superior to analyses based solely on the integrated counting signal. This happens, for instance, when the ancilla is driven through the cavity mode, in which case the presence of the ion affects the temporal fluctuations but not the mean intensity, reflected from the cavity. We also study incoherent processes and ways to reduce their detrimental effects on the readout process.

We proceed to show how two remote qubits ions, each coupled to their own readout ancilla ions, can be brought into a maximally entangled state in a probabilistic manner.
As illustrated in Fig.~\ref{F1}(b), the proposal involves mixing of the output fields from the two cavities through a beam-splitter and detecting the mixed signal with efficient photon counters. Similar proposals have employed the correlation between the photon polarization and Zeeman sublevels of light emitting ions~\cite{Crocker:19}, while our scheme uses different ions for the light emission and the storage of the entangled state.

The article is structured as follows. 
In Sec.~\ref{S2}, we introduce the model, which is followed by a brief overview of the stochastic master equation and conditioned dynamics in Sec.~\ref{S3}.
In Sec.~\ref{S4}, we introduce the Bayesian inference method and its application for qubit readout for different cases in Sec.~\ref{Sec4A}, ~\ref{Sec4B} and ~\ref{sec:coherentDriving}. 
In Sec.~\ref{S5}, we extend the architecture to multiple qubits and show that a continuous measurement of the emitted signal can generate maximally entangled states between remotely situated qubits. 
Finally, we conclude in Sec.~\ref{S6}.

\section{Model}\label{S2}
We consider the system depicted in Fig.~\ref{F1}(a), where two different rare-earth ions with known, distinct transition frequencies are spectrally separate from all other ions in the crystal.
Since rare-earth ions in crystals suffer from inhomogeneous broadening, a small crystal will display distinct and well separated transition frequencies.
The \textit{qubit} ion has an excited state $\ket{e}$ and two long-lived ground states $\ket{0}$ and $\ket{1}$,  which define the logical qubit.
We assume this ion to be far detuned from the cavity resonance.
A second \textit{readout} ion (ancilla) of a different species has only two states $\ket{\downarrow}$ and $\ket{\uparrow}$ and is coupled resonantly to the cavity mode. 
For simplicity, we assume the bad cavity limit such that the cavity mode may be adiabatically eliminated, resulting in an effective Purcell enhanced decay rate $\gamma$ of the readout ion.
If the two ions are in the vicinity of each other, they experience a dipole-dipole interaction~(\ref{eq:dipole}), with a strength
\begin{equation}
\mu = \Big(\frac{\epsilon+2}{3\epsilon}\Big)^2 \frac{\mu_q\mu_r}{4\pi\epsilon_0 r^3_m}[\pmb{\hat{\mu}}_r\cdot\pmb{\hat{\mu}}_q - 3(\pmb{\hat{\mu}}_r\cdot\pmb{\hat{r}}_m)(\pmb{\hat{\mu}}_q\cdot\pmb{\hat{r}}_m)],   
\end{equation}
where $\vec{r}_m$ is the vector between their positions, $\pmb{\mu}_{r(q)}= \mu_{r(q)}\pmb{\hat{\mu}}_{r(q)}$ is the difference in the permanent dipole moment between the excited and the ground state of the readout(qubit) ion and
$\epsilon$ is the relative permittivity at zero frequency, taking into account the local field corrections due to the crystal host material.
For typical values of dipole moments, this energy shift is in the order of $1$ GHz, $1$ MHz and $1$ kHz for an ion separation of $1$ nm, $10$ nm and $100$ nm, respectively~\cite{OHLSSON200271,PhysRevB.46.5912}. 

The dipole interaction~(\ref{eq:dipole}) imposes a frequency shift of the readout ion conditioned on the state of the qubit ion. 
Hence, the fluorescence signal obtained when the readout ion is subjected to a classical drive is conditioned on the state of the qubit ion. 
This mechanism effectively conveys information about the state of the qubit ion.

The excited state $\ket{e}$ is crucial to facilitate the coupling between the two ions, but it is more prone to pertubations than the ground states and it   has a non-negligible decay rate $\Gamma$ (for simplicity, we shall assume an even branching to the two ground states $\ket{0}$ and $\ket{1}$).
Our qubit is hence stored and processed in the ground states $\ket{0}$ and $\ket{1}$, and only for the readout protocol, we apply a $\pi$ pulse on the $\ket{0}\leftrightarrow\ket{e}$ to temporarily transfer the population in $\ket{0}$  to $\ket{e}$ while the  $\ket{1}$ population is unchanged.

We are now in a position to infer the original state of the qubit by analyzing the excitation dynamics of the readout ion. We first investigate the case where the readout ion is driven directly with a laser field with a Rabi frequency $\Omega_{\uparrow\downarrow}$, entering, e.g, from a direction perpendicular to the cavity mode, so that only light scattered by the ion may leave through the cavity mirror. In the frame rotating at the laser frequency, this leads to a Hamiltonian of the form
\begin{align}
H_{\rm readout} = -\delta \ket{\uparrow}\bra{\uparrow} 
+ 
\frac{\Omega_{\uparrow\downarrow}}{2}\Big(\ket{\downarrow}\bra{\uparrow} +\ket{\uparrow}\bra{\downarrow} \Big),
\label{E3}
\end{align}
where $\delta$ is the atom-field detuning which we set to zero in the following.

We then proceed in Sec.~\ref{sec:coherentDriving} to study a more practical model where the cavity is driven by a coherent field of amplitude $\beta$. 
In this way, the fluorescence intensity from the ion is mixed with the reflection of the driving field on the cavity mirror into the detection channel. If no excitation is lost to other channels, all incident photons are reflected, and it is by the temporal correlations rather than by the integrated photon signal, that we shall be able to infer the state of the qubit ion.  

\section{Photon counting and conditional dynamics}\label{S3}
The continuous driving of the readout ion results in a fluorescence signal which is detected using a photon counter.
During any given short time interval $dt$, the photon counter has the possibility to detect either one photon or none.
A photon detection is accompanied by a quantum jump of the emitter to its ground state and occurs in a time interval $dt$ with a probability of $dp= {\rm Tr}[\hat{C}_r\rho(t)\hat{C}^{\dagger}_r]dt$, where $\hat{C_r} = \sqrt{\gamma}(\ket{\downarrow}\bra{\uparrow})$, and $\gamma$ is the Purcell enhanced decay rate of the ancilla ion due to coupling to the rapidly decaying (and adiabatically eliminated) cavity mode. 
Upon detection of a photon, the density matrix $\rho$ of the combined system of the qubit and readout ion is updated according to  
\begin{align}
\rho(t) \xrightarrow[\text{jump}]{\text{}} \hat{C}_r\rho(t)\hat{C}^{\dagger}_r,    
\label{eq:P1}
\end{align}
while in the absence of any photon detection, with probability $1-dp$, $\rho$ propagates according to the no-jump master equation-
\begin{align}
\dot{\rho}_{\rm no\;jump}= -i[H, \rho] -\frac{1}{2} \{\hat{C}^{\dagger}_r\hat{C}_r, \rho\}.   
\label{eq:P2}
\end{align}
In either case the density matrix is subsequently renormalized by the factor $1/dp$ or $1/(1-dp)$, respectively.   
$H$ denotes the Hamiltonian of the qubit system.
We define a variable $dN_t$, which takes values $1$ or $0$, depending on whether the detector registers a photon or not, and the complete detection record, consisting of such \textit{click} and \textit{no-click} events, from $t= 0$ to $T$, is denoted $\{dN_t\}_{t = 0}^T$. 
When averaged over a large number of independent realizations of the photo current, the stochastic dynamics comply with the Lindblad master equation for the ensemble averaged dynamics, which differs from \eqref{eq:P2} by an additional sandwich term, $\hat{C}_r\rho(t)\hat{C}^{\dagger}_r$. Below, we shall add extra terms to this equation to represent the unobserved decay of the qubit system.

\section{Bayesian inference for qubit readout}\label{S4}
As described above, the system dynamics during a time interval $[0,T]$ is conditioned on the specific realization of the measurement signal $\{dN_t\}_{t = 0}^T$, which is, by Eqs.~(\ref{eq:P1}) and ~(\ref{eq:P2}), governed by the evolution of the quantum state. While the total count $N_T$ accumulated until the final time $T$ holds some information about the initial qubit state, the temporal correlations in the full sequence $\{dN_t\}_{t = 0}^T$  may contribute significant further sensitivity to the physical parameters governing the dynamics and the initial qubit state  ~\cite{PhysRevA.92.032124, PhysRevA.97.052113, PhysRevA.95.022306}.

This information may be extracted by Bayes' theorem which updates the prior probabilities  $P(h_i)$ of given hypotheses ($h_1, h_2, h_3,...$), conditioned on a specific measurement outcome on the system. In our case this measurement outcome is the entire detection record $\{dN_t\}_{t = 0}^T$, and Bayes' rule states 
\begin{equation}\label{eq:bayes}
P(h_i|\{dN_t\}_{0}^T)= \frac{P(\{dN_t\}_{0}^T|h_i)P(h_i)}{\sum_j\, P(\{dN_{t}\}_{0}^T|h_j)P(h_j)},
\end{equation}
where $P(\{dN_t\}_{t = 0}^T|h_i)$ is the probability to obtain the record $\{dN_t\}_{t=0}^T$ if hypothesis $h_i$ is true. While it may seem a formidable task to calculate these probabilities and to evaluate the sum in the denominator for any such record, the information is in fact already at hand. The quantum jump sequence corresponding to the record $\{dN_t\}_{t=0}^T$, occurs precisely with the probabilities $dp$ and $1-dp$ for the jumps and the no-jump intervals that we listed above. For $P(\{dN_t\}_{0}^T|h_i)$, this yields a simple product of the jump and no-jump probabilities encountered along the evaluation of the conditional quantum states $\rho_i$, which we initiate and propagate separately according to the different hypotheses. We are not interested in the probability of a given, actually observed record relative to other unobserved ones, and we are allowed to renormalize the probabilities as long as we retain the ratio between the different candidate  hypotheses. The sum in the denominator of \eqref{eq:bayes} merely normalizes the probability distribution of the hypotheses.

As more time is allocated for detection, more information (clicks and intervals with no clicks) becomes available and the probabilities assigned by Bayes rule~(\ref{eq:bayes}) converge.
For our task, we consider two hypotheses, which are the possible initial states of the qubit. That is  $h_0:\, \rho_0 =  \ket{0}\bra{0}\otimes \ket{\downarrow}\bra{\downarrow}$ and $h_1:\, \rho_1 = \ket{1}\bra{1}\otimes \ket{\downarrow}\bra{\downarrow}$. 
At the final time $T$, the hypothesis $h_j$ with the largest probability assigned by Bayes rule~(\ref{eq:bayes}) is chosen as the outcome of the state measurement.

We characterize the achievements of the  Bayesian inference scheme by the probability of assigning a \textit{false} hypothesis upon obtaining a given time series of photo detect events,
\begin{align}\label{E7}
\begin{split}
Q_E(T) &= P({\rm choose}\, h_0|h_1)P(h_1) 
\\
& + P({\rm choose}\, h_1|h_0)P(h_0).
\end{split}
\end{align}

For $\mu\gg \Omega$, the readout ion is tuned completely
out of resonance from the driving field if the qubit is prepared in $\ket{0}$. 
Hence, no fluorescence occurs and the detection of just a single photon signifies that the qubit is in $\ket{1}$.
The error probability~(\ref{E7}) is in this case determined by the probability of having no clicks given that the qubit is in $\ket{1}$,
\begin{align}\label{eq:QeLargeMu}
Q_E(T) = P({\rm no\,clicks\,for\,t\in[0,T]}|\psi = \ket{1}) P(h_1).
\end{align}
The probability $P({\rm no\,clicks\,for\,t\in[0,T]}|\psi = \ket{1})$ is the trace of the no-jump density matrix initialized in  $\rho_1= \ket{1}\bra{1}\otimes \ket{\downarrow}\bra{\downarrow}$ and propagated subject to the no-jump master equation~(\ref{eq:P2}),
and we find the analytic expression
\begin{align}\label{eq:QeLargeMuAnalytic}
Q_E(T) = 
\frac{4\Omega_{\uparrow\downarrow}^2 - \gamma^2\cos(\tilde{\Omega} t) + 2\gamma\tilde{\Omega}\sin(\tilde{\Omega}t)}{4\tilde{\Omega}^2}
P(h_1)\e{-\gamma t/2}
,
\end{align}
with $\tilde{\Omega} = \sqrt{\Omega_{\uparrow\downarrow}^2-\gamma^2/4}$.
We note that the error probability vanishes for large times $t\gg \gamma^{-1}$, as it becomes exponentially unlikely to have no emission events from the resonantly driven readout ion.  
To evaluate the performance of our proposal for intermediate values of $\mu$, and in the presence of  unmonitored decoherence channels, we have recourse to numerical simulations of the stochastic dynamics.
While in an experiment, the detection record $\{dN_t\}_{t = 0}^T$ is delivered by the photo counter, to find the average probabilities  $P({\rm choose }\, h_i|h_j)$, 
we here simulate a large number $\mathcal{N} =20,000$ of such records. For each of these \textit{trajectories}, Bayes rule is applied to evaluate the $P(h_j|\{dN_{t}\}_{t = 0}^{T})$ with $j = 0,1$ and the most likely hypothesis is identified. 
This is repeated for each of the two possible initial states $\ket{0}$ and $\ket{1}$, yielding
$P({\rm choose }\, h_i|h_j)$ from the number of occurrences where the initial state was $\rho_j$, and the given trajectory favoured the hypothesis $h_i$.

For comparison, we consider also the error probability if the state discrimination is based on the integrated signal $N_T$. 
This is the conventional analysis of many experiments but it neglects the information held by the temporal correlations in the counting signal, and it is expected to have a lower performance than the full Bayesian analysis. 
The distributions of the total count $P(N_T|h_0)$ and $P(N_T|h_1)$ have different mean values depending on the qubit state. For example, at large $\mu\gg \Omega_{\downarrow\uparrow}$, the total count under hypothesis $h_0$ is $0$, while according to $h_1$, the mean number of clicks should be $\frac{\gamma \Omega_{\downarrow\uparrow}^2 }{\gamma^2+2\Omega_{\downarrow\uparrow}^2}T$.
In the numerical examples, we sample the 
the total count distributions $P(N_T|h_0)$ and $P(N_T|h_1)$ from the 20,000 simulated counting signals.
Assuming an equal prior probability of $1/2$, the average error probability is then given by $1/2$ times $\sum_{N_T} {\rm Min}[P(N_T|h_0),P(N_T|h_1)]$. 

\subsection{Finite dipole coupling strength $\mu$}\label{Sec4A}
\begin{figure*}
\centering
\includegraphics[width=0.98\textwidth]{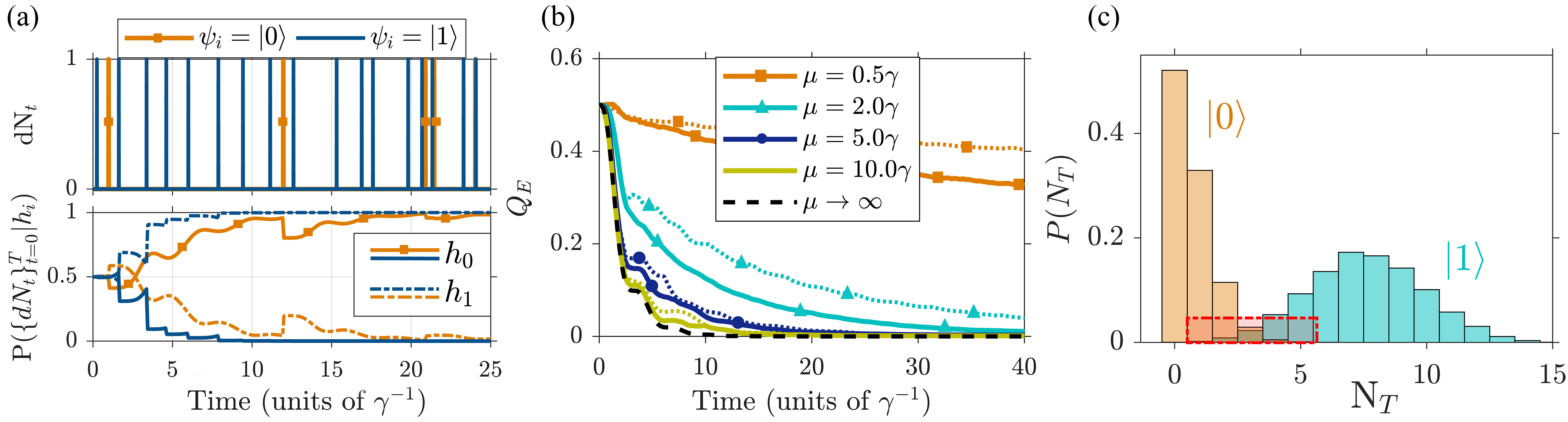}
\caption{Continuous driving and monitoring of the readout ion assuming a long lived qubit excited state $\ket{e}$. 
(a) 
Upper panel:
Simulated measurement records $dN_t$ from $t= 0$ to $t= 25\gamma^{-1}$ for two possible cases: $\psi_i= \ket{0}$ and $\psi_i= \ket{1}$ (blue thin lines) with $\mu= 5\gamma$, $\Omega= 2\gamma$.
Lower panel:
Bayesian probabilities of the two hypotheses $h_0$ and $h_1$ conditioned on the records shown in the upper panel. Bold (dot-dashed)
lines correspond to the hypothesis $h_0$ $(h_1)$.
Orange line with square markers (blue lines) show probabilities conditioned on the counting signals from $\psi_i = \ket{0}$ ($\psi_i = \ket{1}$) in the upper panel.
(b) Error probabilities $Q_E(T)$ for different values of the dipole-dipole interaction $\mu$ found from $\mathcal{N}= 20,000$ simulated detection records. The black, dashed line correspond to the analytical expression when $\mu\gg\gamma$. The dotted lines correspond to the incomplete inference based on the integrated signal $N_T$. 
(c) Normalized probability distributions of the total count $N_T$ for each of the two true states $\ket{0}$ and $\ket{1}$ shown for $\mu= 5\gamma$ at the time $T= 17.40 \gamma^{-1}$. 
}
\label{F2}
\end{figure*}

The top panel of Fig.~\ref{F2}(a) shows typical detection records $dN_t$ registered from a single simulation of Eq.~(\ref{eq:P1}) and ~(\ref{eq:P2}) for each of the two possible initial states assuming the interaction strength,  $\mu = 5\gamma$.
When the qubit is initially in $\ket{1}$, the $\pi$-pulse on the $\ket{0}\leftrightarrow \ket{e}$ transition leaves the qubit state unchanged and the readout ion exhibits usual Rabi oscillations with frequent photon counts as shown by the blue record.
On the other hand, when the qubit is initially in $\ket{0}$, the $\pi$-pulse transfers the population to $\ket{e}$.
This activates the dipole-dipole interaction between the two ions and shifts the energy level of the readout ion by an amount depending on $\mu$, leading to a reduced number of detection events as shown by the orange record.

The Bayesian analysis considers two hypotheses, $h_i$ ($i= 0,1$), which are the possible initial states of the qubit.
The likelihood of each hypothesis conditioned on the detection records $\{dN_t\}^T_{t=0}$ shown in the upper panel, given by $P(h_i|\{dN_t\}^T_{t=0})$ are shown in the lower panel of Fig.~\ref{F2}(a).
In both cases, the hypotheses are assigned equal prior probabilities, $P(h_0) = P(h_1) = 1/2$, and we see that periods with no detected photons ($dN_t= 0$) lead to a smooth evolution of the Bayesian probabilities, favouring the hypothesis $\rho_0$. This is intuitive since the absence of photons consolidates the belief that the readout ion is in the ground state.
For the same reason, discrete jumps, favouring $\rho_1$, occur at each photon detection ($dN_t= 1$) until the probabilities converge to the true hypothesis at the final time. 
It is interesting to note that when the qubit is in $\ket{1}$, the ancilla ion exhibits Rabi oscillations which govern the emission probability and this explains the oscillations between two click events in the lower panel of Fig.~\ref{F2}(a).

The evolution of the error probability $Q_E(T)$ is shown in Fig.~\ref{F2}(b) for different strengths of the dipole coupling $\mu$.
For all values, we observe a convergence to $Q_E(T) = 0$ as $T\rightarrow \infty$. However, the error probability decreases more slowly for weaker interactions $\mu$, where the conditional energy shift of $\ket{\uparrow}$ is less pronounced, allowing almost equally frequent photon detections under both hypotheses $h_0$ and $h_1$.
For larger $\mu$, an appreciable difference in the detection record (frequent detections under $h_1$ and no detections under $h_0$) occur, resulting in fast and reliable inference of the correct state. We see that the numerical results, indeed, approach the analytic expression~(\ref{eq:QeLargeMuAnalytic}) in the limit $\mu\gg\gamma$.

In Fig.~\ref{F2}(c), we exemplify the distributions of the total count $N_T$ accumulated during a time $T= 17.4 \gamma^{-1}$ (a duration chosen here to yield an illustrative histogram) for $\mu= 5\gamma$. While it is clear that the qubit state $\ket{0}$ facilitates, on average, fewer photon emissions than the state $\ket{1}$, a finite overlap between the two distributions still persists. The associated error probabilities are shown as dotted lines in Fig.~\ref{F2}(b) where we see that the integrated signal delivers a larger error in the readout than the full signal treated in a Bayesian analysis.
The advantage of the full signal is more pronounced for smaller values of $\mu$ where the blockade of the readout ion is far from complete, such that the two qubit states allow more similar fluorescence signal.

\subsection{Effects of decay of the qubit excited state $\ket{e}$}\label{Sec4B}
We have seen that under ideal settings and with sufficient time available, it is possible to perfectly infer the qubit state from the measurement signal. However, any qubit system will suffer from some dissipative coupling to its environment. In the Introduction we proposed candidate rare-earth ion systems with orders of magnitude difference in their excited state lifetimes, but to illustrate the effects of dissipation more clearly, and to describe cases with less favorable parameters, we shall here consider qubit excited state decay rates $\Gamma$, just one or two orders of magnitude smaller than the readout ion decay rate $\gamma$. Such parameter regimes include the possibility to use ions of the same species, where the qubit ion is detuned away from the cavity resonance and does not experience the Purcell enhanced decay rate of the resonant readout ion.
For simplicity, we consider equal decay rates $\Gamma$  to each of the two qubit ground states $\ket{0}$ and $\ket{1}$, and we retain the jump dynamics \eqref{eq:P1} while supplementing the no-jump dynamics of Eq.~\ref{eq:P2} with the additional Lindblad terms,
$-\frac{1}{2}\sum_{n=0,1} \{\hat{C}_n^{\dagger}\hat{C}_n,\rho\} + \hat{C}_n\rho \hat{C}_n^{\dagger}$ , with the unmonitored qubit decay operators  $\hat{C}_n=\sqrt{\Gamma}(\ket{n}\bra{e})$ (with $n=0,1$).

\begin{figure}
\centering
\includegraphics[width=0.45\textwidth]{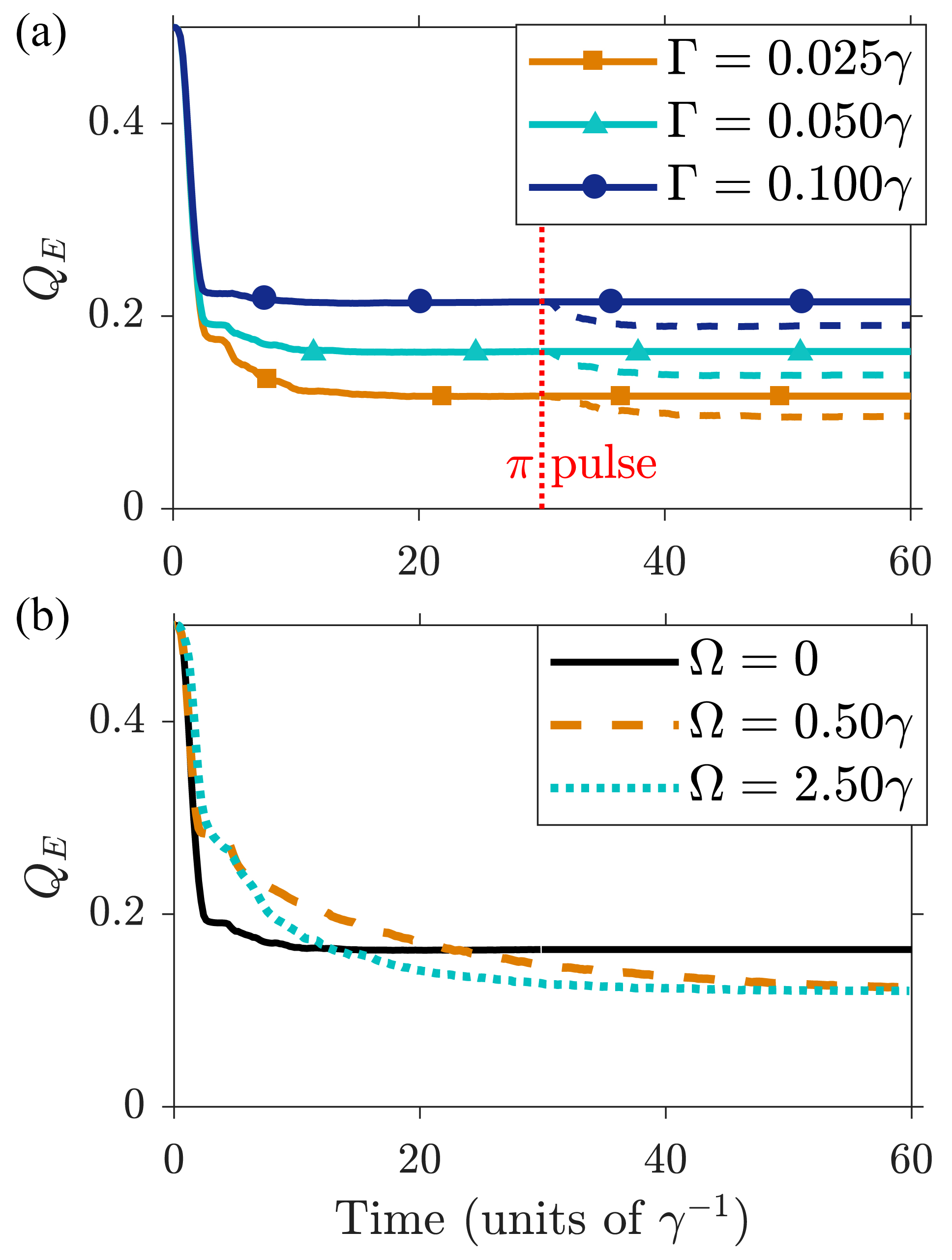}
\caption{Bayesian inference in the presence of excited state decay. (a) Error probability $Q_E(T)$ for different values of $\Gamma$. The dashed lines correspond to the case when a $\pi$-pulse is applied to the $\ket{0}\leftrightarrow\ket{e}$ transition at the time $t_{\pi}= 30\gamma^{-1}$. (b) $Q_E(T)$ for $\Gamma= 0.05\gamma$ when the qubit-ion is subjected to a continuous driving of the $\ket{0}\leftrightarrow\ket{e}$ transition with Rabi frequency $\Omega$. 
Results are shown for $\mu= 5\gamma$ and $\mathcal{N}= 20,000$ and for different values of $\Omega$.}
\label{F3}
\end{figure}

The simulations proceed as above, and we sample the error probability $Q_E(T)$ from $20,000$ simulated detection records $\{dN_t\}_{t=0}^{T}$ subject to the Baysian analysis.
The results are shown as full lines with markers in Fig.~\ref{F3}(a) for $\mu= 5\gamma$ and different values of $\Gamma$.

For the initial qubit state $\ket{0}$ where the initializing $\pi$-pulse would bring the qubit ion to $\ket{e}$, the decay process disengages the dipole coupling as the ion decays into $\ket{0}$ or $\ket{1}$.
This mechanism eventually renders the photo current indifferent to the initial state of the qubit ion, resulting in an error probability $Q_E(T)$ which saturates at a non-zero value around $0.2$ for the values of $\Gamma$ considered here. This effect is enhanced with larger decay rates $\Gamma$.

The saturation occurs once the qubit ion excited state has with certainty decayed to a statistical mixture  of the two ground states $\rho_{\rm qubit} = (\ket{0}\bra{0}+\ket{1}\bra{1})/2$.
By reexciting the ion from the $\ket{0}$ state component, it is possible to extract further information, as half of the population undergoing the decay returned to its original ground state $\ket{0}$ while the other half has become indistinguishable from the other hypothesis, i.e., the initial state $\ket{1}$.
As seen from the dotted lines in Fig.~\ref{F3}(a), a $\pi$ pulse excitation applied at the time $t_{\pi}= 30\gamma^{-1}$ allows a further reduction in the error probability by approximately 15\%.
To extract further information, one can apply multiple $\pi$-pulses until all the population has been transferred to the $\ket{1}$ state.

Another possibility to improve the error probability in the presence of dissipation is to apply a continuous Rabi drive $\Omega$ on the $\ket{0}\leftrightarrow\ket{e}$ transition throughout the readout process as described by the Hamiltonian term
\begin{equation}
H_{\rm drive}= \frac{\Omega}{2}\Big(\ket{0}\bra{e} + \ket{e}\bra{0}\Big).
\label{E9}
\end{equation}

The resulting error probabilities for two different driving strengths $\Omega = 0.50\gamma$ and $\Omega = 2.50\gamma$
are displayed in Fig.~\ref{F3}(b). We note that at short times $t\lesssim\Gamma^{-1}$, the driving decreases our ability to discern the two hypotheses compared to the case of no driving. This is because the Rabi oscillation between $\ket{e}$ to $\ket{0}$ periodically disengages
the dipole-dipole interaction. At longer times, however, the drive serves its purpose to reexcite the qubit and thereby restores some sensitivity to the qubit initial state.
The error probability nonetheless saturates at a final value as in the pulsed scheme of Fig.~\ref{F3}(a), allowing for these parameters a lowering of $Q_E(T)$ by approximately $26$\%.
The saturation at a finite $Q_E$ for long times is due to the inevitable shelving of the population in the uncoupled qubit state $\ket{1}$, producing indistinguishable photo currents.

\subsection{Inference by light reflected from the cavity}\label{sec:coherentDriving}
In the previous subsections, we considered the case when the readout ion was directly driven with a Rabi frequency $\Omega_{\uparrow\downarrow}$. 
Experimentally, it is convenient to drive the readout ion by a pump field incident on the cavity. The readout signal shown in Fig.~\ref{F1}(a) then contains both the reflected laser field from the cavity mirror and the signal transmitted through the mirror from inside the cavity.
If there are no internal losses and the readout ion decays only by the Purcell enhanced cavity emission, all incident photons are eventually detected and the total number of detection events carry no information about the qubit state. The same is not true for the signal record $\{dN_t\}^T_{t=0}$ which carries temporal correlations due the interaction with the readout ion. The optimal inference protocol is also for this situation described by the quantum trajectory formalism and Bayes rule.

The cavity mode is continuously driven by a coherent source of amplitude $\beta$, and to formally describe this situation we assume a weak Jaynes-Cummings coupling constant $g$ of the read-out ion to the cavity and a rapid cavity decay rate $\kappa$ through the input mirror, which permits adiabatic elimination of the quantum state of the cavity field. The intracavity field will be of magnitude $2\beta/\sqrt{\kappa}$, while a correction to the field due to the interaction with the read-out ion will act back on the dipole and cause its Purcell enhanced damping with a rate $\gamma=4g^2/\kappa$. We ignore other decay channels for the readout ion and if we assume no internal cavity losses, input-output theory yields an output field described by the interference of the intra-cavity field, expressed as a sum of $2\beta/\sqrt{\kappa}$ and a readout ion contribution, and the reflected driving field. The result is an output field given as $\hat{C}_r+\beta$. Since this plays the role of the annihilation operator of the output detected field, it is convenient to rewrite Eq.(\ref{eq:P1}) and (\ref{eq:P2})  with $\hat{C}_r+\beta$ appearing instead of $\hat{C}_r$. This rewriting, in turn, causes a correction to the Hamiltonian, and yields the net interaction in Eq.(\ref{E3}), with the value of $\Omega_{\uparrow\downarrow}/2 = g\beta/\sqrt{\kappa}=\beta\sqrt{\gamma}/2$.

The new equations describe quantum jumps due to the detection of a photon in the reflected field, and these jumps now occur with a mean probability given by the classical flux of photons $|\beta|^2$ while individual jumps have  a weaker back action on the quantum state of the ions due to the c-number component of the jump operator. . 

\begin{figure}
\centering
\includegraphics[width=0.45\textwidth]{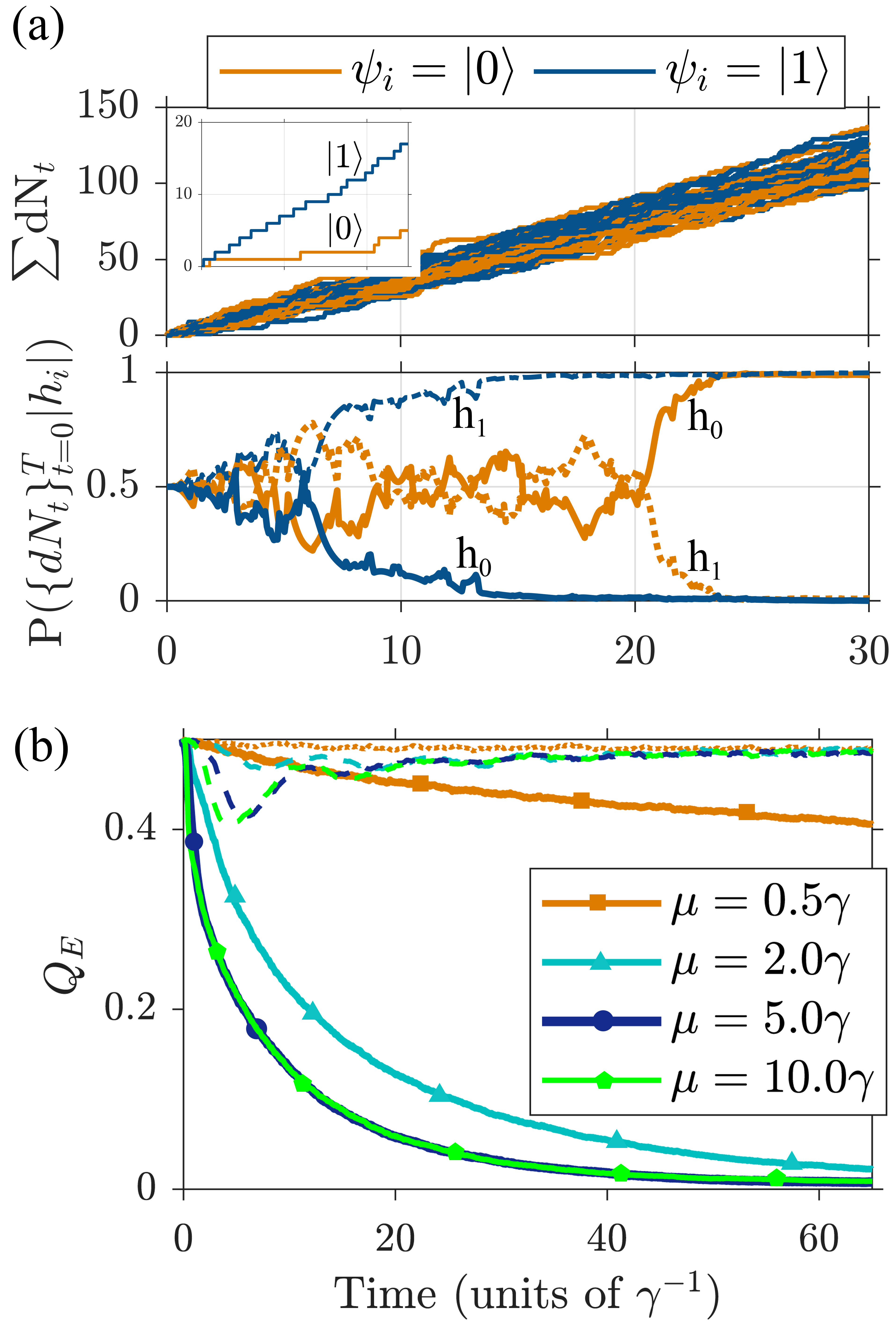}
\caption{Continuous monitoring of the ancilla using photon counting on the signal reflected from the cavity, driven by a coherent field of amplitude $\beta$. (a) Upper panel: Cumulative sum of the measurement record $\sum_{t=0}^{T} dN_t$ for 20 realizations of the two possible cases of $\psi_i$. The insert shows $\sum_{t=0}^T dN_t$ for the case of direct driving studied in Sec.~\ref{Sec4A}.
Lower panel: 
Bayesian probabilities of the two hypotheses $h_0$ and $h_1$ conditioned on two of the records shown in the upper panel. Bold (dot-dashed)
lines correspond to the hypothesis $h_0$ $(h_1)$.
Orange (blue) lines show probabilities conditioned on one of the counting signals from $\psi_i = \ket{0}$ ($\psi_i = \ket{1}$) in the upper panel.
(b) Average inference error probability $Q_E(T)$ as a function of time for different values of dipole-dipole interaction $\mu$ sampled from $\mathcal{N}= 20,000$ realizations. The dotted line corresponds to inference based on the total number of detection events. In all the simulations, $\beta^2= 4\gamma$.
}
\label{F4}
\end{figure}

In the top panel of Fig.~\ref{F4}(a), we plot the cumulative sum of the detection events $\sum_{t=0}^T dN_t$ as a function of time for $20$ realizations simulated using the above replacements in Eq.~\ref{eq:P1} and Eq.~\ref{eq:P2}.  While the active read-out ion does not change the total number of counts it modifies the fluctuations in the reflected signal. Due to the strong reflected component of the probe field, we are not in a regime where we observe the anti-bunching of the bare ion fluorescence, but rather in a regime where the large classical $\beta$ amplitude serves as a local oscillator for homodyne detection of the ion signal~\cite{wiseman_milburn_2009}. The power spectrum of the homodyne signal current could reveal the emission components by the ion, while the trajectory (Bayesian) analysis registers correlation to all orders and is even more sensitive to the ion signal contribution, see, e.g., Ref.~\cite{PhysRevA.94.032103}.

The strength of the coherent source $\beta$ is maintained at $2\gamma$,    
and we collect around $30|\beta|^2 = 120$ photons during the time $30\gamma^{-1}$.
However, as envisaged above, the total count does not statistically differ between the two hypotheses, as evident from Fig.~\ref{F4}(a), where orange and blue curves correspond to $\psi_i= \ket{0}$ and $\psi_i= \ket{1}$ respectively.
For comparison, $\sum \{dN_t\}^T_{t=0}$, is shown in the inset of panel (a) for the case of direct driving.

In the lower panel of Fig.~\ref{F4}(a), we show examples of the conditioned probabilities for the two hypotheses $h_i$ ($i= 0,1$), where the bold (dotted) curves correspond to the correct (incorrect) hypothesis.
It is interesting to observe that although the total number of detected photons are indistinguishable  for the two cases in the long time limit, the probabilities of the two hypotheses conditioned on the full measurement signal $\{dN_t\}^T_{t=0}$ can efficiently infer the state of the qubit. 
Contrary to Fig.~\ref{F2}(a) where each click triggers a jump in the probabilities, we now see a smoother, more continuous convergence since each of the numerous photo detections, likely stemming from the driving source, provides much less information. 

In Fig.~\ref{F4}(b), we plot $Q_E(T)$ as a function of time for different values of $\mu$. 
For short times, the initial transient dynamics of the readout ion implies a transient temporal dependence of the cummulated count $N_T$ on the qubit state and the $Q_E$ (dashed lines) decreases correspondingly. However, for longer probing times, the integrated signal $N_t$ is dominated by the steady state flux which carries no information and the error probability saturates at $0.5$.
The Bayesian analysis (bold curves with markers), on the other hand, infers the qubit states perfectly from the qubit dependent temporal correlations in the counting signal.
Absolute discrimination is reached faster for larger values of $\mu$ but it may be noted that in general the inference time is longer compared to the case when the readout ion is driven directly in Fig.~\ref{F2}(b).

While the Bayesian analysis of the full measurement record thus turned out to be crucial to properly infer the qubit state, we note that extraction of average statistical correlations from the count record beyond the mean detection rate may also provide a direct quantitative criterion to infer the qubit state.   
For brevity, we omit a detailed investigation of the effects of dissipation of the qubit ion for this model. The detrimental effects of a decaying excited state are similar to those observed in Sec.~\ref{Sec4B}, and 
a continuous Rabi drive $\Omega$ on the qubit $\ket{0}\leftrightarrow\ket{e}$ transition can improve the error probability $Q_E$ but only until the ion has become shelved in the state $\ket{1}$.

\section{Entanglement of remote qubits}\label{S5}
Quantum entanglement is a precious and crucial resource in many quantum protocols, and any relevant quantum computing platform must be able to produce entangled states between qubits. In the following, we discuss how the ancilla based architecture sketched in Fig.~\ref{F1}(a) can be used in a probabilistic entanglement generation scheme for two remote rare-earth ion qubits. Our proposal is based on ideas developed in Ref.~\cite{PhysRevA.85.032327, PhysRevLett.83.5158, PhysRevLett.89.237901, PhysRevLett.91.067901, PhysRevLett.91.097905} for a wide range of systems.
As illustrated in Fig.~\ref{F1}(b), we consider two identical cavities $A$ and $B$, each containing a qubit ion with long-lived states $\ket{0}_{(A/B)}$ and $\ket{1}_{(A/B)}$, and a excited states $\ket{e}_{(A/B)}$
which are coupled to the readout ion via the dipole-dipole interaction~(\ref{eq:dipole}). 
The readout ions, with energy levels $\ket{\uparrow}_{(A/B)}$ and $\ket{\downarrow}_{(A/B)}$, have a short lifetime and are subjected to a continuous drive with Rabi frequencies $\Omega^A_{\uparrow\downarrow}$ and $\Omega^B_{\uparrow\downarrow}$, where the superscript ${(A/B)}$ represents the two cavities.
Instead of monitoring the emission from individual cavities which appear in modes $\hat{a}_1$ and $\hat{a}_2$, we combine the two fluorescence signals in a 50:50 beam splitter and continuously monitor the mixed signals $\hat{a}_1 + \hat{a}_2$ and $\hat{a}_1 - \hat{a}_2$ in the output ports using two single photon detectors. Since photon detectors are insensitive to frequency and phase, photons arriving from each of the cavities are indistinguishable such that detections effectively entangle the sources $A$ and $B$. 

In our model, where the cavity and traveling light fields are eliminated, this measurement procedure causes a backaction on the two-qubit system represented by the two jump operators, $\hat{\chi}_+ \equiv \frac{1}{\sqrt{2}}(\hat{C}^A_r+\hat{C}^B_r)$ and $\hat{\chi}_-\equiv\frac{1}{\sqrt{2}}(\hat{C}^A_r-\hat{C}^B_r)$, 
where $\hat{C}^{(A/B)}_r = \sqrt{\gamma}\ket{\downarrow}_{(A/B)}\bra{\uparrow}_{(A/B)}$.
Considering a photon counter of quantum efficiency $\eta$, such that $0\leq\eta\leq 1$, the state of the system upon detection of a photon due to a quantum jump in $\hat{\chi}_+$ or $\hat{\chi}_-$ leads to update of $\rho$ as follows-
\begin{align}
\rho(t) \xrightarrow[\text{jump}]{\text{}} \eta \hat{\chi}_{\pm}\rho(t)\hat{\chi}^{\dagger}_{\pm}
\label{eq:SME21}
\end{align}
In the absence of any photon detection, the no-detected-jump master equation is given by
\begin{equation}
\begin{array}{lll}
\dot{\rho}_{\rm no\;jump}= && -i[H,\rho] + \sum_{\alpha=\pm}\Big[(1-\eta)\hat{\chi}_{\alpha}\rho\hat{\chi}^{\dagger}_{\alpha} \\
&&- \frac{1}{2}\{\hat{\chi}^{\dagger}_{\alpha}\hat{\chi}_{\alpha}, \rho\}\Big].
\label{eq:SME22}
\end{array}
\end{equation}
Since the photon detector has an efficiency of $\eta$, this implies that in a fraction of $(1-\eta)$ times, emitted photons pass undetected. 
Such undetected jump events are incorporated in the term with prefactor $(1-\eta)$ in the equation above.

To generate maximally entangled states between the qubits, both qubits are initialized in a superposition state $\frac{1}{\sqrt{2}}(\ket{0} + \ket{1})$. 
This is followed by a $\pi$ pulse on the $\ket{0}\rightarrow\ket{e}$ transition, which engages the dipole-dipole interaction between $\ket{e}$ and $\ket{\uparrow}$ in both cavities. 
We simulate an experiment by solving the stochastic master equation~(\ref{eq:SME21}), (\ref{eq:SME22}), and at the final time a $\pi$-pulse is performed on the $\ket{e}\rightarrow\ket{0}$ transition in order to restore the ions to the qubit subspace. 

\begin{figure}
\centering
\includegraphics[width=0.45\textwidth]{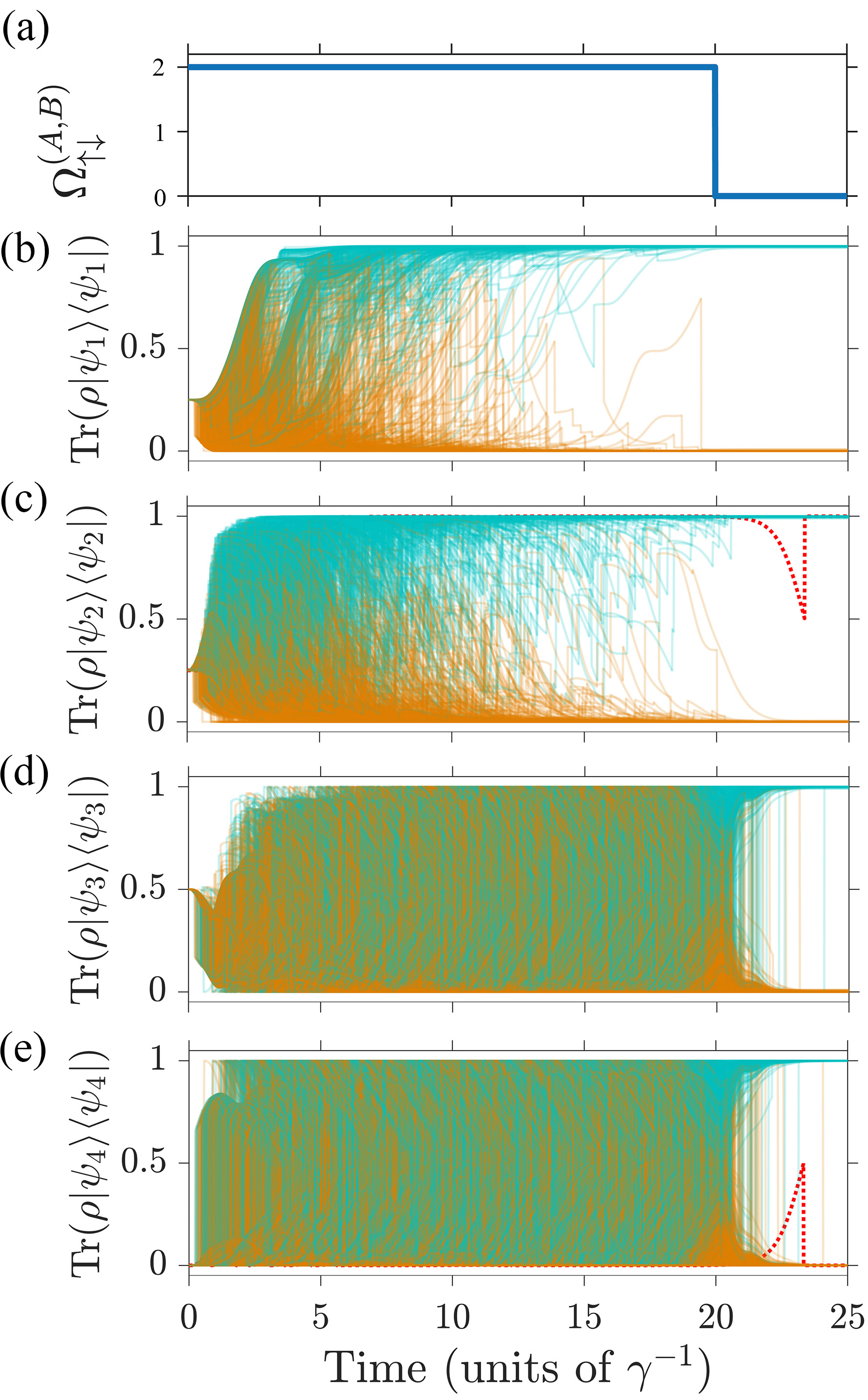}
\caption{Entanglement between remote qubits by continuous photon counting of the mixed signals from the two cavities. Panel (a) shows the Rabi drive $\Omega^{(A,B)}_{\uparrow\downarrow}$ as a function of time. 
Panels (b-e) show the population in each of the states, $\ket{\psi_i}$, given by $|C_i|^2$ subjected to continuous monitoring of the mixed signal where  $i= 1,2,3,4$ (b,c,d,e) for $2500$ runs. 
The trajectories reaching a value $\mathrm{Tr}(\rho \ket{\psi_i}\bra{\psi_i})>0.99 (<0.01)$ are indicated with cyan (orange) in each panel.
}
\label{F5}
\end{figure}

One caveat of this scheme is that it produces an entangled state between the qubits and the readout ions, such that upon tracing out the readout ions, the qubits are left in a mixed state. We propose to eliminate this issue by
turning off the Rabi drives of the readout ions for a duration $t\gg\gamma^{-1}$ at the end of the protocol as shown in Fig.~\ref{F5}(a).
The readout ions then decay to their ground states and factor out, and as long as the emission is monitored (with $\eta = 1$), the purity of the  qubit state is ensured.
The qubit state can at this point be written
\begin{align}
\ket{\Psi} = \sum_{i = 1}^{4}\alpha_i \ket{\psi_i},
\end{align}
where we define the two-qubit basis
\begin{align}
    \ket{\psi_1}&= \ket{0,0}_{(A,B)}, \nonumber \\
    \ket{\psi_2}&= \ket{1,1}_{(A,B)}, \nonumber \\
    \ket{\psi_3}&= \frac{1}{\sqrt{2}}(\ket{0,1} + \ket{1,0})_{(A,B)}, \nonumber \\
    \ket{\psi_4}&= \frac{1}{\sqrt{2}}(\ket{0,1} - \ket{1,0})_{(A,B)}.
\end{align}
Here $\ket{\psi_1}$ and $\ket{\psi_2}$ are (undesirable) product states, while $\ket{\psi_3}$ and $\ket{\psi_4}$ are maximally entangled states between the two qubits.

The scheme  can be intuitively understood as follows. The initial state is a superposition with 25\% population in each of the product states $\ket{\psi_1}$, and $\ket{\psi_2}$ and 50\% population in the entangled state $\ket{\psi_3}$. 
After excitation by a $\pi$ pulse from $\ket{0}$ to the excited state $\ket{e}$ , the qubits control the ancilla emission. 
The measurement scheme involves counting ancilla photons, and we thus effectively measure the total occupation of the qubit excited state, which can have- (i) high flux from two emitting ancilla ions, (ii) average flux from one emitting ancilla or (iii) low flux  when emission from both ancilla ions is suppressed. (i) and (iii) correspond to the qubit product states $\ket{\psi_{1}}$ and $\ket{\psi_2}$, while, due to the beam splitter, the photon count rate corresponding to precisely one emitting ancilla ion does not distinguish which cavity emits. Hence it yields a projection on $\ket{\psi_{3}}$ or $\ket{\psi_{4}}$. The relative sign between the state components of the entangled states changes by each count event in the output arm that carries the ``minus” component of the interfering signal, so the two entangled states are also distinguished by the measurement sequences, and for long probing times they will heralded by the detection record with equal $25$\% probabilities.

In Fig.~\ref{F5}(b-e), we display the state populations $\mathrm{Tr}(\rho \ket{\psi_i}\bra{\psi_i})$ as functions of time. Each panel contains results of 2,500 independent realizations. We observe that in every run, the state has been projected on exactly one of the states $\ket{\psi_i}$ at the final time (i.e. $\mathrm{Tr}(\rho \ket{\psi_i}\bra{\psi_i})=1$ for any one value of $i$, while the others are 0). 
In each panel, trajectories reaching a value $\mathrm{Tr}(\rho \ket{\psi_i}\bra{\psi_i})>0.99$ are indicated with cyan, while the remaining are orange.
From $\mathcal{N}= 20,000$  independent simulations, we observe that the four states occur with equal probability.
This implies that this protocol has a $50$\% chance of heralding a maximally entangled state. 

As shown in Fig.~\ref{F5}(b-e) the collapse to $\ket{\psi_1}$ and $\ket{\psi_2}$ in (b) and (c) occurs faster than to the two entangled states in (d) and (e). This is because the number of emitted photons takes distinctly lower and higher values for the two product states than for the two entangled states while the population switches between the two entangled state at each detection associated with the jump operator $\hat{\chi}_-$. Such an  event may be triggered by decay of a readout ion even after the resonant drive is turned off at $t=20\gamma^{-1}$, and the qubit state only chooses between $\ket{\psi_3}$ and $\ket{\psi_4}$ according to which last jump operator $\hat{\chi}_\pm$ applies.
In the simulations, we notice a single  event around $t = 23\gamma^{-1}$ where one of the trajectories exhibits a very unlikely behaviour (marked with dotted red curve in Fig.~\ref{F5}(c)~and~(e)). 
This event is caused by the qubits predominantly occupying the product state $\ket{\psi_2}$, but then a rare long period with no photo detection in one of the detectors causes a rotation towards the entangled state $\ket{\psi_3}$. This rotation would have continued if a detection event had not abruptly projected the state  back into $\ket{\psi_2}$.

Our results demonstrate that under ideal conditions, the measurement-based scheme is able to produce a high fidelity entangled pair of remote qubits with a heralding probability of $50\%$. Due to the insensitivity of counters to photon phases, the protocol is robust to \textit{known} phase and frequency variations between the two sources. However, under uncontrolled phase fluctuations   between the two components A and B, it will not be possible to distinguish $\ket{\psi_3}$ and $\ket{\psi_4}$ by the measurement signal, and the scheme fails. 

\begin{figure*}
\centering
\includegraphics[width=0.9\textwidth]{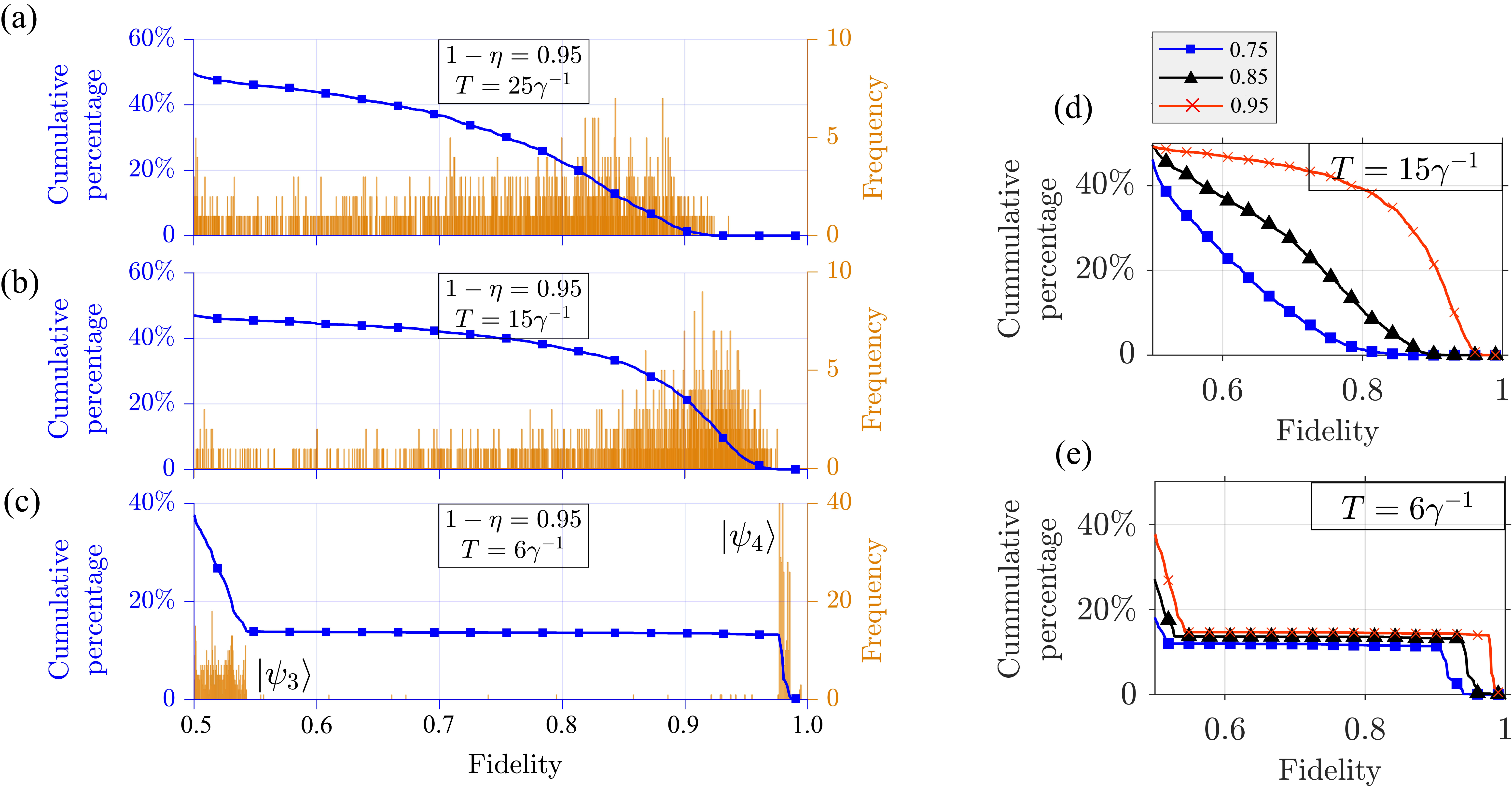}
\caption{Heralding of high fidelity entangled states with a finite detector efficiency .
The orange histograms  show the distribution of $\ket{\psi_3}$ and $\ket{\psi_4}$ fidelities for $\mathcal{N}= 2500$ independent measurement trajectories, and the blue curves with square markers illustrate the cumulative percentage of trajectories surpassing different values of the fidelity above $0.5$. The detector efficiency is $(1-\eta)= 0.95$, and the probing time assumes the values $T= 25 \gamma^{-1}$ in panel (a) ,  $T= 15 \gamma^{-1}$ in panel (b), and $T= 6\gamma^{-1}$ in panel (c). The vast difference between state $\ket{\psi_3}$ and $\ket{\psi_4}$ fidelities in panel (c) is explained in the text.
 Panels (d) and (e) show the cumulative percentage of trajectories surpassing different values of the fidelity for different detector efficiencies $(1-\eta)=0.95,\ 0.85$ and $0.75$ (from above).
}
\label{F6}
\end{figure*}

Even though state-of-the art photon detectors can reach very high efficiencies, propagation loss and finite photon detection efficiency imply that a finite fraction $\eta$ of the photons are undetected as represented by the term with the prefactor $(1-\eta)$ in the stochastic master equation~(\ref{eq:SME22}). 
For finite values of $\eta$, the trajectories which would ideally lead to the maximally entangled states $\ket{\psi_{3}}$ and $\ket{\psi_{4}}$ now lead to mixed states. There is a trade-off between too short probing times which do not adequately distinguish $\ket{\psi_{3}}$ and $\ket{\psi_{4}}$ from $\ket{\psi_{1}}$ and $\ket{\psi_{2}}$ and too long probing times where the increased probability of missing a photon detection event translates directly into uncertainty about the sign in the entangled state superposition. We study this trade-off, exploiting the fact that for a given detection record $dN_t$, the stochastic master equation yields a definite final mixed state. Based on its entangled state content, we may choose to reject or retain the outcome, and thus exchange success probability for an increase in fidelity of the heralded state. 

In Fig.~\ref{F6}, we present the distribution of fidelities of any one of the two maximally entangled states $\ket{\psi_{3}}$ and $\ket{\psi_{4}}$ as extracted from $\mathcal{N}= 2500$ independent simulations of the experiment with a detector efficiency of $(1-\eta)=0.95$ and different durations of the experiment. The orange histograms reveal a variety of values of the fidelities, and we recall that since the fidelity is known in each run of the experiment, it is possible to discard low-fidelity events and retain only a smaller fraction of events with a higher fidelity. The blue curve with markers show the cumulative percentage of the runs that exceed the fidelity argument on the $x$-axis of the plots. In panel (a), about $5\%$ of trajectories reach fidelities above $0.9$ when $(1-\eta)= 0.95$. The absence of trajectories reaching near-unit fidelities in Fig.~\ref{F6} (a) is due to the long probing time of $T= 25 \gamma^{-1}$ and the resulting high probability of undetected photons. Indeed, the shorter probing time $T= 15 \gamma^{-1}$ in Fig.~\ref{F6} (b) leads to a visibly larger fraction of trajectories with high fidelities. In Fig.~\ref{F6} (c) we present results for the even shorter probing time  $T= 6\gamma^{-1}$, and here we observe a clear division between experimental runs leading to heralding of a high fidelity  $\ket{\psi_{4}}$ state, and runs that signal a dominant $\ket{\psi_{3}}$ component with a much lower fidelity. The difference between the two is caused by the conservation of the initial exchange symmetry of the state under no-click evolution and clicks governed by the jump operator $\hat{\chi}_+$, which only exchanges the population between the symmetric states $\ket{\psi_{1}}$, $\ket{\psi_{2}}$ and $\ket{\psi_{3}}$, while a single, early application of the jump operator $\hat{\chi}_-$ suffices to significantly populate the initially unoccupied anti-symmetric state $\ket{\psi_{4}}$.
In Fig.~\ref{F6} (d) and (e) we show the dependence of the cumulative percentage of trials that surpass different values of the fidelity for three different values of $(1-\eta)= 0.95,\ 0.85$ and $0.75$. 
When probed for  $T= 15\gamma^{-1}$ (panel (d)), the number of high fidelity outcomes drops drastically as the efficiency of the detector decreases. 
For the shorter probing time $T= 6\gamma^{-1}$, (panel (e)), however, the scheme performs better due to the symmetry arguments explained above, and one may achieve the entangled state $\ket{\psi_4}$ with fidelities above $0.9$ even with the single-photon detector efficiency as low as $75\%$.

\section{Conclusions}\label{S6}
We have investigated the possibility of using an ancilla ion to read out the state of a qubit rare-earth ion in crystals.  
The ions exhibit large state-dependent dipole-moments, leading to dipole-dipole interaction between them when they are situated in close vicinity.
We propose to exploit this interaction to infer the state of the qubit ion by monitoring the emission from a continuously driven ancilla, whose dynamics depend on the state of the qubit.
This requires that two species of closely lying ions are spectroscopically identified in the crystal, one serving as a qubit with excellent coherence and lifetime properties and another as an ancilla with a closed two-level transition and stronger coupling to its optical surroundings.
Using a stochastic master equation, we simulated experiments and showed how a Bayesian analysis of the measurement signal obtained from photon detection can infer the initial state of the qubit.
When the readout ion is driven via the cavity input mirror, as e.g., in experiments with fiber cavities, a Bayesian analysis extracts temporal correlations in the counting signals crucial for the distinction of the qubit states.

Next, we showed how the same architecture may be used to probabilistically create maximally entangled states between remote qubits.
To this end, we considered a system with two cavities, each with its own qubit, which is dipole-coupled to a continuously driven ancilla. We showed that by mixing the two fluorescence signals on a beamsplitter whose output ports are continuously monitored, the two-qubit state can be stochastically collapsed to an entangled state with $50\%$ success probability.
This scheme, in its simplest form, suffers from propagation and detection losses, but we demonstrated that one may optimize the process duration and retain fewer outcome states and thus exchange success probability for fidelity of the heralded state.  

Our analysis took its starting point in previous proposals for ancilla-mediated read out by fluoresence detection. We assumed a weak coupling of the bad cavity mode to the ancilla ion, while in the regime of strong coupling and a good cavity, it is possible to detect the qubit state populations by their ability to split and shift the cavity resonance~\cite{rempe_nature}. Similar schemes can be implemented with our ancilla ion, benefiting from the possibility to separately optimize the qubit properties of one ion and the optical properties of the readout ion. The readout would here be done by a phase shift measurement, and sequential illumination of multiple cavities could be used to entangle qubits with a higher tolerance to detector inefficiency~\cite{PhysRevA.98.030302}.          

\section{Acknowledgements}
The authors acknowledge support from the European Union FETFLAG program, Grant No. 820391 (SQUARE) and from the Danish National Research Foundation through the Center of Excellence ”CCQ” (Grant agreement no.:DNRF156). 

\bibstyle{apsrev4}

\newpage

\end{document}